\documentclass[twocolumn,english,aps,prx,showpacs]{revtex4-1}

\usepackage[T1]{fontenc}
\usepackage[latin9]{inputenc}
\usepackage{color}
\usepackage{babel}
\usepackage{amsmath}
\usepackage{amssymb}
\usepackage{graphicx}
\usepackage[unicode=true,pdfusetitle,
 bookmarks=true,bookmarksnumbered=false,bookmarksopen=false,
 breaklinks=false,pdfborder={0 0 1},backref=false,colorlinks=true]
 {hyperref}

\makeatletter
\usepackage{babel}
\usepackage{babel}
\usepackage{babel}
\usepackage{babel}
\usepackage{babel}
\usepackage{babel}

\makeatother

\begin{document}

\title{Unitary dynamics of strongly-interacting Bose gases \\
 with time-dependent variational Monte Carlo in continuous space}

\author{Giuseppe Carleo}

\affiliation{Institute for Theoretical Physics, ETH Zurich, Wolfgang-Pauli-Str. 27,
8093 Zurich, Switzerland }

\author{Lorenzo Cevolani}

\affiliation{Laboratoire Charles Fabry, Institut d'Optique, CNRS, Univ. Paris
Sud 11, 2 avenue Augustin Fresnel, F-91127 Palaiseau cedex, France}

\author{Laurent Sanchez-Palencia}

\affiliation{Laboratoire Charles Fabry, Institut d'Optique, CNRS, Univ. Paris
Sud 11, 2 avenue Augustin Fresnel, F-91127 Palaiseau cedex, France}

\affiliation{Centre de Physique Théorique, Ecole Polytechnique, CNRS, Univ Paris-Saclay,
F-91128 Palaiseau, France }

\author{Markus Holzmann}

\affiliation{LPMMC, UMR 5493 of CNRS, Universit{é} Grenoble Alpes, F-38042 Grenoble,
France}

\affiliation{Institut Laue-Langevin, BP 156, F-38042 Grenoble Cedex 9, France}
\begin{abstract}
We introduce time-dependent variational Monte Carlo for continuous-space
Bose gases. Our approach is based on the systematic expansion of the
many-body wave-function in terms of multi-body correlations and is
essentially exact up to adaptive truncation. The method is benchmarked
by comparison to exact Bethe-ansatz or existing numerical results
for the integrable Lieb-Liniger model. We first show that the many-body
wave-function achieves high precision for ground-state properties,
including energy and first-order as well as second-order correlation
functions. Then, we study the out-of-equilibrium, unitary dynamics
induced by a quantum quench in the interaction strength. Our time-dependent
variational Monte Carlo results are benchmarked by comparison to exact
Bethe ansatz results available for a small number of particles, and
also compared to quench action results available for non-interacting
initial states. Moreover, our approach allows us to study large particle
numbers and general quench protocols, previously inaccessible beyond
the mean-field level. Our results suggest that it is possible to find
correlated initial states for which the long-term dynamics of local
density fluctuations is close to the predictions of a simple Boltzmann
ensemble. 
\end{abstract}
\maketitle

\section{Introduction}

The study of equilibration and thermalization properties of complex
many-body systems is of fundamental interest for many areas of physics
and natural sciences~\cite{NaturePhysicsInsight2015}. For systems
governed by classical physics, an exact solution of Newton's equations
of motion is often numerically feasible, using for instance molecular-dynamics
simulations. For quantum systems, the mathematical structure of the
time-dependent Schrödinger equation is instead fundamentally more
involved. Quantum Monte Carlo algorithms, the \emph{de facto} tool
for simulating quantum many-body systems at thermal equilibrium \cite{ceperley_path-integrals_1995,0034-4885-75-9-094501,PhysRevA.84.061606},
cannot be directly used to study time-dependent unitary dynamics.
Out-of-equilibrium properties are then often treated on the basis
of approximations drastically simplifying the microscopic physics.
Irreversibility is either enforced with an explicit breaking of unitarity,
e.g.\ within the quantum Boltzmann approach, or the dynamics is reduced
to mean-field description using time-dependent Hartree-Fock and Gross-Pitaevskii
approaches. Although these approaches may qualitatively describe thermalization
\cite{PhysRevA.66.013603,Nature539}, their range of validity cannot
be assessed because genuine quantum correlations and entanglement
are ignored.

For specific systems, exact dynamical results can be derived. This
is the case for integrable 1D models, for which Bethe ansatz (BA)
solutions exist \cite{Gaudin}. However, also in this case many open
questions still persist. For example, the exact evaluation of correlation
functions for out-of-equilibrium dynamics is at present an unsolved
problem. As a result, despite important theoretical and experimental
progress \cite{caux2012constructing,caux2013timeevolution,collura2013equilibration,trotzky2012probing,langen2013localemergence,greif2015formation,gring2012relaxation,cominotti2014optimal},
a complete picture of thermalization (or its absence), e.g. based
on general quench protocols \cite{pasquale_calabrese_time_2006},
is still missing.

Numerical methods for strongly-interacting systems face important
challenges as well. Numerical Renormalization Group (NRG) and Density-Matrix
Renormalization Group (DMRG) approaches provide an essentially exact
description of arbitrary 1D lattice systems in-\ and out-of-equilibrium
\cite{white_density_1992,white_real-time_2004,daley_time-dependent_2004,anders_real-time_2005},
but they have less predictive power when applied to continuos-space
systems. On one hand, multi-scale extensions of the DMRG optimization
scheme to the limit of continuous-space lattices \cite{dolfi_multigrid_2012}
are so-far limited to relatively small system sizes \cite{dolfi_minimizing_2015}.
On the other hand, efficient ground-state optimization schemes for
continuous quantum field matrix product states (c-MPS) \cite{verstraete_continuous_2010}
have been introduced only very recently \cite{ganahl_continuous_2016},
and applications to quantum dynamics are still to be realized. A further
formidable challenge is the efficient extension of these approaches
to higher dimensions, which is a fundamentally hard problem.

Another class of methods for strongly-interacting systems is based
on variational Monte Carlo (VMC), combining highly-entangled variational
states with robust stochastic optimization schemes \cite{PhysRevLett.98.110201}.
Such approaches have been successfully applied to the description
of continuous quantum systems, in any dimension and not only in 1D~\cite{PhysRevB.74.104510,PhysRevB.91.115106}.
More recently, out-of-equilibrium dynamics has become accessible with
the extension of these methods to real-time unitary dynamics, within
time-dependent variational Monte Carlo (t-VMC) \cite{carleo_localization_2012,giuseppe_carleo_light-cone_2014}.
So far, the t-VMC approach has been developed for \textit{lattice}
systems with bosonic \cite{carleo_localization_2012,giuseppe_carleo_light-cone_2014},
spin \cite{cevolani_protected_2015,blass2016absence,carleo2016solving},
and fermionic \cite{PhysRevB.92.245106} statistics, yielding a description
of dynamical properties with an accuracy often comparable with MPS-based
approaches.

In this Paper, we extend t-VMC to access dynamical properties of interacting
quantum systems in continuous space. Our approach is based on a systematic
expansion of the wave function in terms of few-body Jastrow correlation
functions. Using the 1D Lieb-Liniger model as a test case, we first
show that the inclusion of high-order correlations allows us to systematically
approach the exact BA ground state energy. Our results improve by
orders of magnitude on previously published VMC and c-MPS results,
and are in line with latest state-of-the-art developments in the field.
We further compute single-body and pair correlation functions, hardly
accessible by current BA methods. We then calculate the time evolution
of the contact pair correlation function following a quench in the
interaction strength. For the non-interacting initial state, we benchmark
our results to exact BA calculations available for a small number
of bosons and further compare to the quench action approach for large
systems approaching the thermodynamic limit. We finally apply our
method to the study of general quenches from arbitrary initial states,
for which no exact results in the thermodynamic limit are currently
available.

\section{Method}

\subsection{Expansion of the Many-Body Wave-Function}

Consider a non-relativistic quantum system of $N$ identical bosons
in $d$ dimensions, and governed by the first-quantization Hamiltonian
\begin{eqnarray}
\mathcal{H} & = & -\frac{1}{2}\sum_{i=1}^{N}\nabla_{i}^{2}+\sum_{i=1}^{N}v_{1}(\vec{x}_{i})+\frac{1}{2}\sum_{i\neq j}^ {}v_{2}(\vec{x}_{i},\vec{x}_{j}),\label{eq:hamiltonian}
\end{eqnarray}
where $v_{1}(\vec{x})$ and $v_{2}(\vec{x},\vec{y})$ are, respectively,
a one-body external potential and a pair-wise inter-particle interaction~\footnote{We have conveniently set the particle mass and the reduced Planck
constant to unity.}. Without loss of generality, a time-dependent $N$-body state can
be written as $\Phi(\mathbf{X},t)=\exp\left[U(\mathbf{X};t)\right]$,
where $\mathbf{X}=\vec{x}_{1},\vec{x}_{2}\dots\vec{x}_{N}$ is the
ensemble of particle positions and $U$ is a complex-valued function
of the $N$-particle coordinates, $\mathbb{R}^{N\times d}\rightarrow\mathbb{C}$.
Since the Hamiltonian (\ref{eq:hamiltonian}) contains only two-body
interactions, it is expected that an expansion of $U$ in terms of
few-body Jastrow functions containing at most $m$-body terms, rapidly
converges towards the exact solution. Truncating this expansion up
to a certain order $M\leq N$, leads to the Bijl-Dingle-Jastrow-Feenberg
expansion \cite{Bijl,Dingle,jastrow_many-body_1955,feenberg_theory_1969}
\begin{multline}
U^{(M)}(\mathbf{X};t)=\sum_{i=1}u_{1}(\vec{x}_{i};t)+\frac{1}{2!}\sum_{i\neq j}u_{2}(\vec{x}_{i},\vec{x}_{j};t)+\\
+\dots\frac{1}{M!}\sum_{i_{1}\neq i_{2}\neq\dots i_{M}}u_{M}(\vec{x}_{i_{1}},\vec{x}_{i_{2}}\dots\vec{x}_{i_{M}};t),\label{eq:JFE}
\end{multline}
where $u_{m}(\mathbf{r};t)$ are functions of $m$ particle coordinates,
$\mathbf{r}=\vec{x}_{i_{1}},\vec{x}_{i_{2}}\dots\vec{x}_{i_{m}}$,
and of the time $t$. A global constraint on the function $U(\mathbf{X},t)$
is given by particle statistics. In the bosonic case, we demand that
$U(\vec{x}_{1},\vec{x}_{2}\dots\vec{x}_{N})=U(\vec{x}_{\sigma(1)},\vec{x}_{\sigma(2)}\dots\vec{x}_{\sigma(N)})$,
for all particle permutations $\sigma$. In general, the functions
$u_{m}(\mathbf{r};t)$ can have an arbitrarily complex dependence
on the $m$ particle coordinates, which can prove problematic for
practical applications. Nonetheless, a simplified functional dependence
can often be imposed, resulting from the two-body character of the
interactions in the original Hamiltonian. For $m\geq3$, $u_{m}(\mathbf{r};t)$
can be conveniently factorized in terms of general two-particle vector
and tensor functions following Ref.~\cite{PhysRevB.74.104510}. Details
of this approach and the present implementation are presented in Appendix~\ref{sec:Functional-Structure-of}
and \ref{sec:3body}.

An appealing property of the many-body expansion (\ref{eq:JFE}) is
that it is able to describe intrinsically non-local correlations in
space. For instance the two-body $u_{2}(\vec{x}_{i},\vec{x}_{j};t)$,
as well as any $m$-body function $u_{m}(\mathbf{r};t)$, can be long
range in the particle separation $\left|\vec{x}_{i}-\vec{x}_{j}\right|$.
This non-local spatial structure allows for a correct description
of gapless phases, where a two-body expansion may already capture
all the universal features, in the sense of the renormalization group
approach \cite{kane_general_1991}. This is in contrast with the MPS
decomposition of the wave-function, which is intrinsically local in
space.

\subsection{Time-Dependent Variational Monte Carlo}

The time evolution of the variational state \eqref{eq:JFE} is entirely
determined by the time-dependence of the Jastrow functions $u_{m}(\mathbf{r};t)$.
In order to establish optimal equations of motion for the variational
parameters, we start noticing that the functional derivative of $U(\mathbf{X};t)$
with respect to the variational, complex-valued, Jastrow functions
$u_{m}(\mathbf{r};t)$, 
\begin{align}
\frac{\delta U(\mathbf{X};t)}{\delta u_{m}(\mathbf{r};t)} & \equiv\rho_{m}(\mathbf{r}),\label{eq:funcderu}
\end{align}
yields the $m$-body density operators 
\begin{align}
\rho_{m}(\mathbf{r}) & =\frac{1}{m!}\sum_{i_{1}\neq i_{2}\neq\dots i_{m}}\prod_{j}\delta(x_{i_{j}}-r_{j}).\label{eq:rhom}
\end{align}
The expectation values of the operators $\rho_{m}$ over the state
$\vert\Phi(t)\rangle$ give the instantaneous $m$-body correlations.
For instance, $\langle\rho_{2}(r_{1},r_{2})\rangle_{t}=\sum_{i<j}\langle\delta(x_{i}-r_{1})\delta(x_{j}-r_{2})\rangle_{t}$,
where $\left\langle \dots\right\rangle _{t}=\left\langle \Phi(t)\right\vert \dots{}\left\vert \Phi(t)\right\rangle /\left\langle \Phi(t)\vert\Phi(t)\right\rangle $,
is proportional to the two-point density-density correlation function.

We can then express the time derivative of the truncated variational
state $U^{(M)}(\mathbf{X};t)$ using the functional derivatives, Eq.~(\ref{eq:funcderu}),
as a sum of the few-body density operators up to the truncation $M$,
i.e. 
\begin{align}
\partial_{t}U^{(M)}(\mathbf{X},t) & =\sum_{m=1}^{M}\int d\mathbf{r}\rho_{m}(\mathbf{r})\partial_{t}u_{m}(\mathbf{r};t).\label{eq:dtum}
\end{align}
The exact wave-function satisfies the Schr{ö}dinger equation $i\partial_{t}U(\mathbf{X};t)=E_{\text{loc}}(\mathbf{X},t)$,
where $E_{\text{loc}}(\mathbf{X};t)=\frac{\left\langle \mathbf{X}\right|\mathcal{H}\left|\Phi(t)\right\rangle }{\left\langle \mathbf{X}\right|\left.\Phi(t)\right\rangle }$
is the so-called \emph{local energy}. The optimal time evolution of
the truncated Bijl-Dingle-Jastrow-Feenberg expansion~(\ref{eq:JFE})
can be derived imposing the Dirac-Frenkel time-dependent variational
principle \cite{dirac_note_1930,frenkel_wave_1934}. In geometrical
terms, this amounts to minimizing the Hilbert-space norm of the residuals
$R^{(M)}(\mathbf{X};t)\equiv\left\Vert i\partial_{t}U^{(M)}(\mathbf{X};t)-E_{\text{loc}}^{(M)}(\mathbf{X};t)\right\Vert $,
thus yielding a variational many-body state as close as possible to
the exact one~\footnote{The natural norm induced by a quantum Hilbert space is the Fubini-Study
norm, which is gauge invariant and therefore insensitive to the unknown
normalizations of the quantum states we are dealing with here.}. The minimization can be performed explicitly and yields a closed
set of integro-differential equation for the Jastrow functions $u_{m}(\mathbf{r};t)$
: 
\begin{eqnarray}
\sum_{p=1}^{M}\int d\mathbf{r}^{\prime}\frac{\delta\left\langle \rho_{m}(\mathbf{r})\right\rangle _{t}}{\delta u_{p}(\mathbf{r}^{\prime};t)}\partial_{t}u_{p}(\mathbf{r}^{\prime};t) & = & -i\frac{\delta\left\langle \mathcal{H}\right\rangle _{t}}{\delta u_{m}(\mathbf{r};t)}.\label{eq:eom}
\end{eqnarray}
In practice, these equations are numerically solved for the time derivatives
$\partial_{t}u_{p}(\mathbf{r}^{\prime};t)$ at each time step $t$.
The expectation values taken over the time-dependent state, $\langle\cdot\rangle_{t}$,
which enter Eq.~\eqref{eq:eom}, are found via a stochastic sampling
of the probability distribution $\Pi(\mathbf{X},t)=\left|\Phi(\mathbf{X},t)\right|^{2}$.
This is efficiently achieved by means of the Metropolis-Hastings algorithm,
as per conventional Monte Carlo schemes (See Appendix~\ref{sec:Monte-Carlo-Sampling}
for details). It then yields the full time evolution of the truncated
Bijl-Dingle-Jastrow-Feenberg state~(\ref{eq:JFE}) after time integration.

The t-VMC approach as formulated here provides, in principle, an exact
description of the real-time dynamics of the $N$-body system. The
essential approximation lies however in the truncation of the Bijl-Dingle-Jastrow-Feenberg
expansion to the $M$ most relevant terms. In practical applications,
the $M=2$ or $M=3$ truncation is often sufficient. Systematic improvement
beyond $M=3$ is possible \citep{PhysRevB.74.104510}, but may require
a substantial computational effort. 
\begin{figure*}[t]
\includegraphics[clip,width=1\columnwidth]{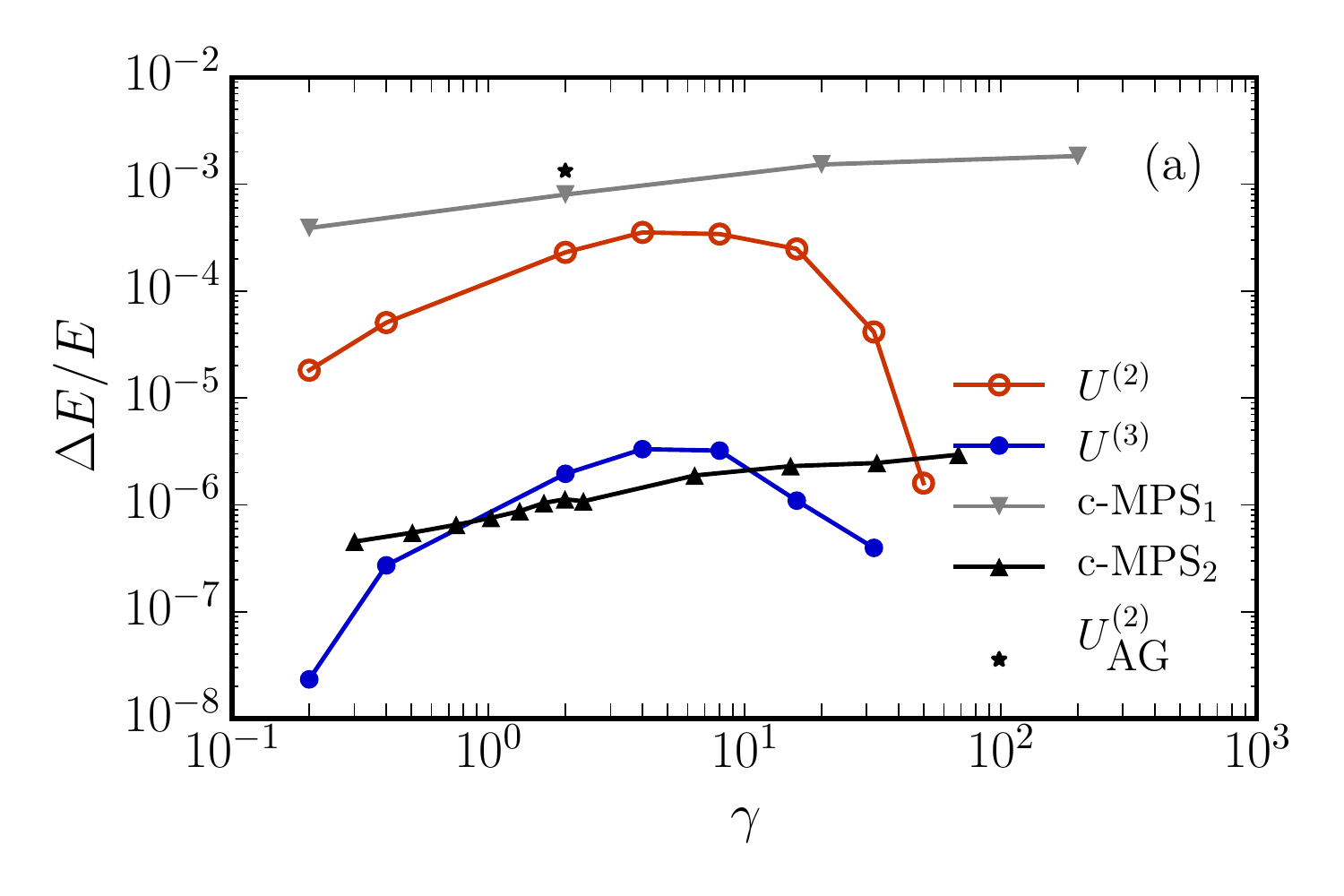}\includegraphics[clip,width=0.5\columnwidth]{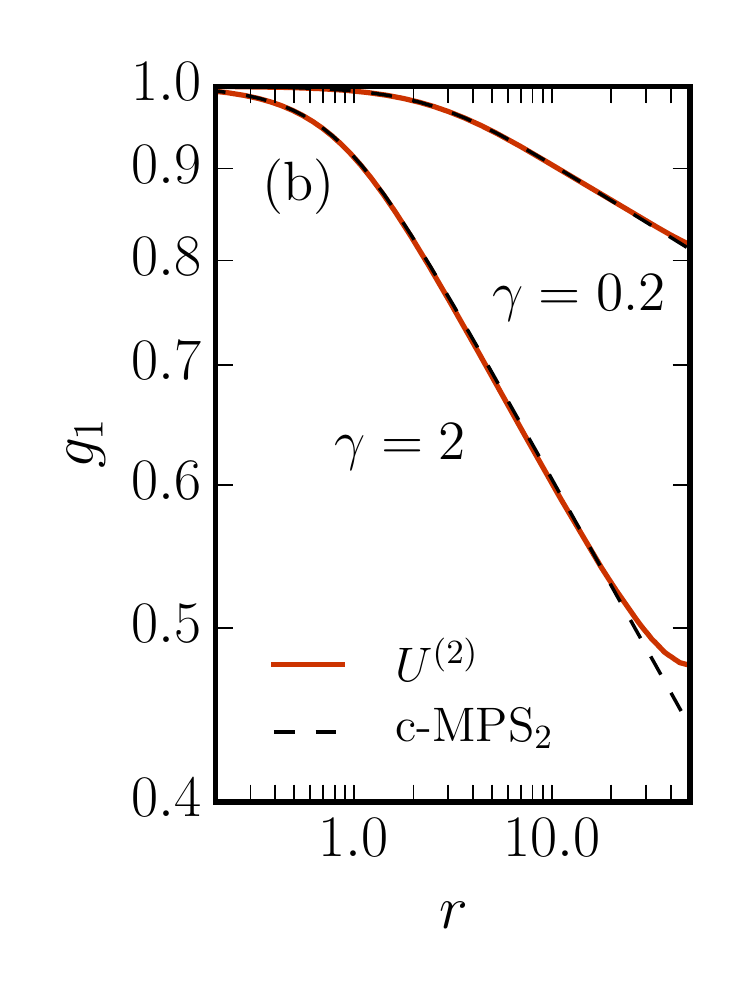}\includegraphics[clip,width=0.5\columnwidth]{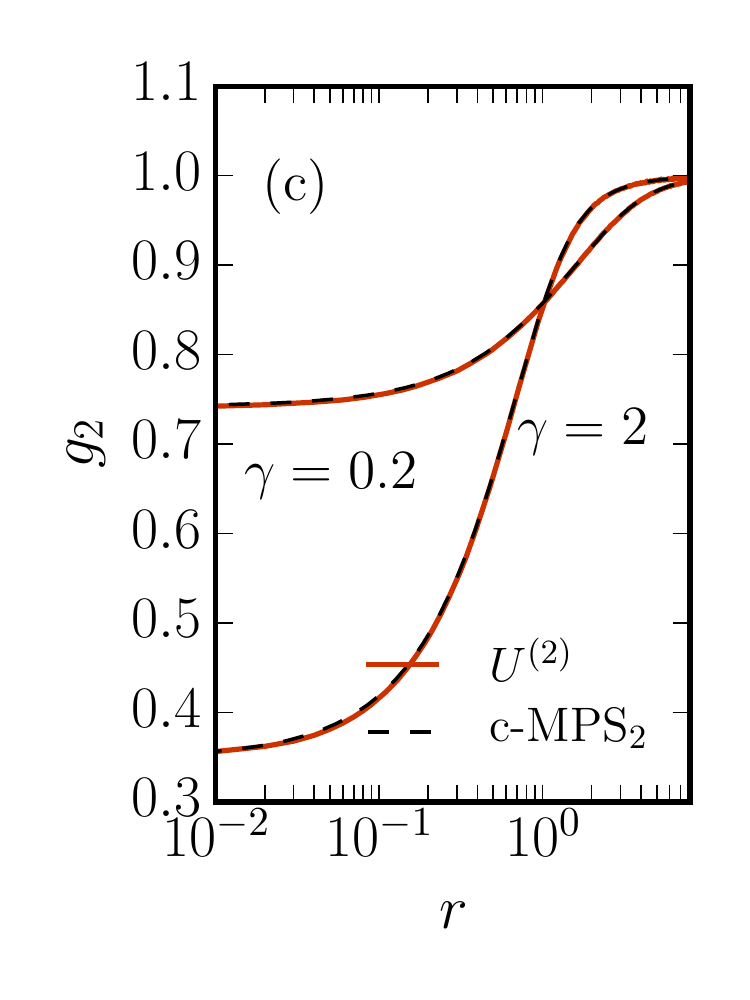}\caption{Ground-state properties of the Lieb-Liniger model as obtained from
different variational approaches : (a) relative accuracy of the ground-state
energy , (b) one-body density matrix, (c) density-density pair correlation
functions. $U^{(2)}$ and $U^{(3)}$ denote results for the 2, and
3-body expansion (present work), $U_{\textrm{AG}}^{(2)}$ is the parametrized
2-body Jastrow state of Ref. \cite{astrakharchik_correlation_2003},
c-MPS$_{1}$ results are from Ref. \cite{verstraete_continuous_2010},
and c-MPS$_{2}$ are those very recently reported in \cite{ganahl_continuous_2016}.
Distances $r$ are in units of the inverse density $1/n$. Our variational
results have been obtained for $N=100$ particles. Finite-size corrections
on local quantities are negligible, and very mildly affect the reported
large-distances correlations. Overall statistical errors are of the
order of symbol sizes, for (a), and line widths, for (b) and (c).
\label{fig:GS-prop}}
\end{figure*}

\section{Lieb-Liniger model}

As a first application of the continuous-space t-VMC approach, we
consider the Lieb-Liniger model~\cite{e._h._lieb_exact_1963}. On
one hand, some exact results and numerical data are available, allowing
us to benchmark the Jastrow expansion and the t-VMC approach. On the
other hand, several aspects of the out-of-equilibrium dynamics of
this model are unknown, which we compute here for the first time using
t-VMC.

The Lieb-Liniger model describes $N$ interacting bosons in one dimension
with contact interactions. It corresponds to the Hamiltonian~(\ref{eq:hamiltonian})
with $v_{1}(x)=0$ and $v_{2}(x,y)=g\delta(x-y)$, where $g$ is the
coupling constant. Here, we consider periodic boundary conditions
over a ring of length $L$, and the particle density $n=N/L$. The
density dependence is as usual expressed in terms of the dimensionless
parameter $\gamma=mg/\hbar^{2}n$. The Lieb-Liniger model is the prototypal
model of continuous one-dimensional strongly-correlated gas exactly
solvable by Bethe ansatz \cite{e._h._lieb_exact_1963}. This model
is experimentally realized in ultracold atomic gases strongly confined
in one-dimensional optical traps, and several studies on out-of-equilibrium
physics have been already realized~\cite{moritz_exciting_2003,kinoshita_quantum_2006,gring2012relaxation,ronzheimer_expansion_2013,fang_quench-induced_2014,boeris_mott_2016}.

As a result of translation invariance, we have $u_{1}(x\:;t)=\text{const}$,
and the first non-trivial term in the many-body expansion~\eqref{eq:JFE}
is the two-body, translation invariant, function $u_{2}(\left|x-y\right|;t)$.
To compute the functional derivatives of the many-body wave function,
we proceed with a projection of the continuous Jastrow fields $u_{m}(\mathbf{r};t)$
onto a finite basis set. Here, we have found convenient to represent
both the field Hamiltonian (\ref{eq:hamiltonian}) and the Jastrow
functions, onto a uniform mesh of spacing $a$, leading to $n_{\text{var}}=\frac{L}{2a}$
variational parameters for the $2$-body Jastrow term. In the following
our results are extrapolated to the continuous limit, corresponding
to $a\rightarrow0$. The finite-basis projection as well as the numerical
time-integration of Eq. (\ref{eq:eom}) are detailed in the Appendix
\ref{sec:Finite-Basis-Projection} and benchmarked against exact diagonalization
results in the three particle case in Appendix~\ref{sec:Lattice-Benchmark}.

\subsection{Ground-state properties}

To assess the quality of truncated Bijl-Dingle-Jastrow-Feenberg expansions
for the Lieb-Liniger model, we start our analysis considering ground-state
properties. An exact solution can be found from the Bethe ansatz (BA)
and gives access to exact ground state energies and local properties
\cite{e._h._lieb_exact_1963}. Other non-local properties are substantially
more difficult to extract from the BA solution, and unbiased results
for ground state correlation functions have not been reported so far.
In order to determine the best possible variational description of
the ground-state within our many-body expansion, different strategies
are possible. A first possibility is to consider the imaginary-time
evolution $\left|\Psi(\tau)\right\rangle =e^{-\tau\mathcal{H}}\left|\Psi_{0}\right\rangle $
which systematically converges to the exact ground-state in the limit
$\tau\gg\Delta_{1}$, where $\Delta_{1}=E_{1}-E_{0}$ is the gap with
the first excited state on a finite system and provided that the trial
state $\Psi_{0}$ is non-orthogonal to the exact ground state. Imaginary-time
evolution in the variational subspace can be implemented considering
the formal substitution $t\rightarrow-i\tau$ in the t-VMC equations
(\ref{eq:eom}). The resulting equations are equivalent to the stochastic
reconfiguration approach \cite{sorella_generalized_2001}. However,
direct minimization of the variational energy can be significantly
more efficient, in particular for systems becoming gapless in the
thermodynamic limit, where $\Delta_{1}\sim\textrm{poly(}1/N)$. Given
the gapless nature of the Lieb-Liniger model, we have found computationally
more efficient to adopt a Newton method to minimize the energy variance
\cite{umrigar_energy_2005}.

For the ground state, the many-body expansion truncated at $M=2$
is exact not only in the non-interacting limit $\gamma=0$ but also
in the Tonks-Girardeau limit $\gamma\rightarrow\infty$. In this fermionized
limit the wave-function can be written as the modulus of a Vandermonde
determinant of plane waves, corresponding to the two-body Jastrow
function $u_{2}(r)=\log\left(\sin r\pi/L\right)$ \cite{girardeau_relationship_1960}.
To assess the overall quality of pair wave functions for ground state
properties, we start comparing the variational ground-state energies
$E$ obtained for $M=2$ with the exact BA result. In Fig.~\ref{fig:GS-prop}(a)
we show the relative error $\Delta E/E$ as a function of in the interaction
strength $\gamma$. We find that the relative error is lower than
$10^{-4}$ for all values of $\gamma$ and the accuracy of our two-body
Jastrow function is superior to previously published variational results
based either on c-MPS \cite{verstraete_continuous_2010} or VMC \cite{astrakharchik_correlation_2003}
{[}see Fig.~\ref{fig:GS-prop}(a) for a quantitative comparison{]}.
Notice that the improvement with respect to previous VMC results is
due to the larger variational freedom of our $u_{2}(r)$ function,
which is not restricted to any specific functional form as done in
Ref.~\cite{astrakharchik_correlation_2003}.

Even though the accuracy reached by the two-body Jastrow function
may already be sufficient for most practical purposes for all values
of $\gamma$, we have also considered higher order terms with $M=3$,
shown in Fig.~\ref{eq:JFE}(a). The introduction of the third-order
term yields a sizable improvement such that the maximum error is about
three orders of magnitude smaller than original c-MPS results \cite{verstraete_continuous_2010},
and feature a similar accuracy of recently reported c-MPS results
\cite{ganahl_continuous_2016}. Overall, our approach reaches a precision
on a continuous-space system, which is comparable to state-of-the-art
MPS/DMRG results for gapless systems on a lattice \cite{pippan2010efficient}.

Finally, to further assess the quality of our ground state ansatz
beyond the total energy, we have also studied non-local properties
of the ground state wave function, which are not accessible by existing
exact BA methods. In Figs.~\ref{fig:GS-prop}(b) and (c) we show,
respectively, our results for the off-diagonal part of the one-body
density matrix, $g_{1}(r)\propto\langle\Psi^{\dagger}(r)\Psi(0)\rangle$,
and for the pair correlation function, $g_{2}(r)\propto\langle\Psi^{\dagger}(r)\Psi(r)\Psi^{\dagger}(0)\Psi(0)\rangle$,
where $\Psi(r)$ is the bosonic field operator. We find an overall
excellent agreement with the results that have been obtained with
c-MPS in Ref. \cite{ganahl_continuous_2016}, except for some small
deviations at large values of $r$ which we attribute to residual
finite-size-effects in our approach. We found that the addition of
the 3-body terms does not change significantly correlation functions.
Already at the 2-body level, the present results are statistically
indistinguishable from exact results obtained using our implementation
\cite{carleo2013universal} of the worm algorithm \cite{boninsegni_worm_2006}
and for the same system (not shown). 
\begin{figure*}[t]
\includegraphics[clip,width=1\columnwidth]{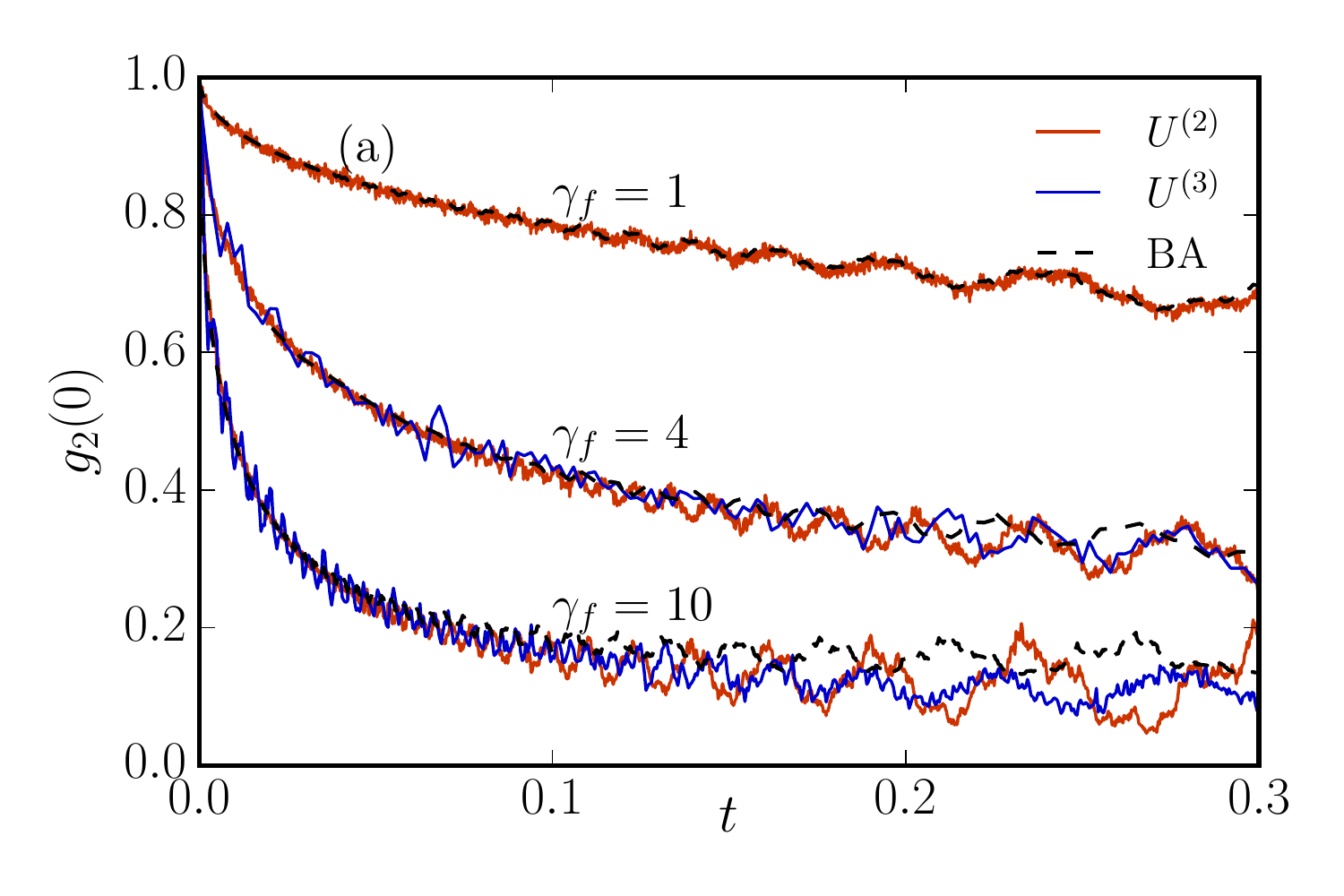}\includegraphics[clip,width=1\columnwidth]{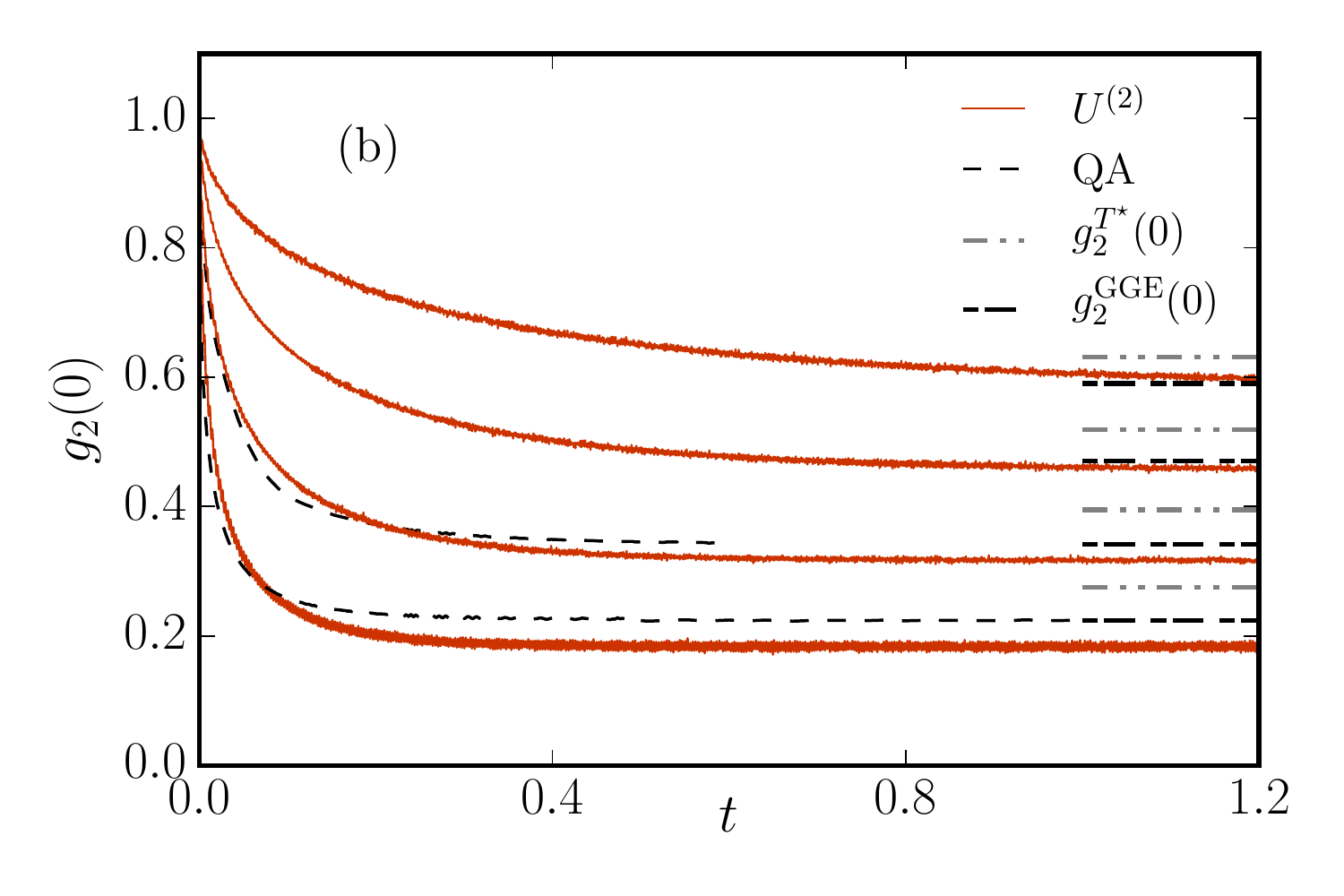}\caption{Time-dependent expectation value of local two-body correlations after
a quantum quench from a non-interacting state, $\gamma_{i}=0$, to
$\gamma_{f}$ : (a) t-VMC results are compared with BA results obtained
for a small number of particles \cite{zill_relaxation_2015,zill_coordinate_2016}.
The correlation function is rescaled to have $g_{2}(0,0)=1$; (b)
t-VMC results for $N=100$ particles and $\gamma_{f}=1,2,4,8$ (from
top to bottom) compared to the quench action predictions from Ref.~\cite{nardis_relaxation_2015}
(dashed lines), to the Boltzmann thermal averages at the effective
temperature $T^{\star}$ (dotted-dashed lines), and to the GGE thermal
averages prediction (rightmost dashed lines). Statistical error bars
on t-VMC data are of the order of lines width. \label{fig:dyn-BA-comp}}
\end{figure*}

\subsection{Quench dynamics}

Having assessed the quality of the ansatz for local and non-local
ground-state properties, we now turn to the study of the out-of-equilibrium
properties of the Lieb-Liniger model. We focus on the description
of the unitary dynamics induced by a global quantum quench of the
interaction strength, from an initial value $\gamma_{i}$ to a final
value $\gamma_{f}$. Exact BA results are available only in the case
of a non-interacting initial state ($\gamma_{i}=0$). Even in this
case, the dynamical BA equations can be exactly solved only for a
modest number of particles with further truncation in the number of
energy eigen-modes \cite{zill_relaxation_2015,nardis_relaxation_2015},
$N\lesssim10$, since the complexity of the BA solution increases
exponentially with the number of particles. Simplifications in the
thermodynamic limit are exploited by the quench action \cite{caux_time_2013},
and have been recently applied to quantum quenches starting from a
non-interacting initial state \cite{nardis_relaxation_2015}. In the
following we first compare our t-VMC results to these existing results,
and then present new results for quenches following a non-vanishing
initial interaction strength.

To assess the quality of the time-dependent wave function we compare
our results for the evolution of local density-density correlations,
$g_{2}(0,t)$, with the truncated BA results obtained in Refs. \cite{zill_relaxation_2015,zill_coordinate_2016}
for a small number of particles, $N\simeq6$. Appendix~\ref{sec:Lattice-Benchmark}
provides also further validation of our method accessing $g_{2}(r,t)$
at non-vanishing distances. The comparison shown in Fig.~\ref{fig:dyn-BA-comp}(a)
shows an overall good agreement. The t-VMC and BA results are indistinguishable
for weak interactions ($\gamma_{f}=1$). For larger interactions,
we notice systematic but small differences between BA and t-VMC with
$M=2$ or $M=3$. These differences amount to a small increase in
the amplitude of the oscillations. This effect tends to increase with
the interaction strength, being hardly visible for $\gamma=1$ and
more pronounced for $\gamma=10$. However, these oscillations result
at large times from the discrete mode structure due to the very small
number of particles. They vanish in the physical thermodynamic limit.
In turn, the comparison at small particle numbers indicates an accuracy
better than a few percent for time-averaged quantities in the asymptotic
large time limit up to $\gamma=10$, with results at $M=3$ systematically
improving on the $M=2$ case. Concerning small and intermediate time
scales, we do not observe systematic deviations between the t-VMC
results and the BA solution. In particular, the relaxation times are
remarkably well captured by the t-VMC approach. On the basis of this
comparison and of the comparison for three particles with exact diagonalization
results presented in Appendix~\ref{sec:Lattice-Benchmark}, we conclude
that t-VMC allows accurate quantitative studies of both the relaxation
and equilibration dynamics. This careful benchmarking now allows us
to confidently apply the t-VMC approach regimes that are inaccessible
to exact BA namely large but finite $N$ and long times, as well as
the case of non-vanishing initial interactions.

Let us consider relaxation of density correlations for a large number
of particles, close to the thermodynamic limit (here we use $N=100$).
As shown in Fig.~\ref{fig:dyn-BA-comp}(b), we notice that the amplitudes
of the large-time oscillations, attributed to the discrete mode spectrum,
are now drastically suppressed compared to the quenches with $N=6$.
After an initial relaxation phase, the quantity $g_{2}(0,t)$ approaches
a stationary value. Comparing our curves with those obtained with
the quench action method, we find a qualitatively good agreement,
albeit a general tendency to underestimate the quench action predictions
is observed.

\begin{figure}[t]
\includegraphics[clip,width=1\columnwidth]{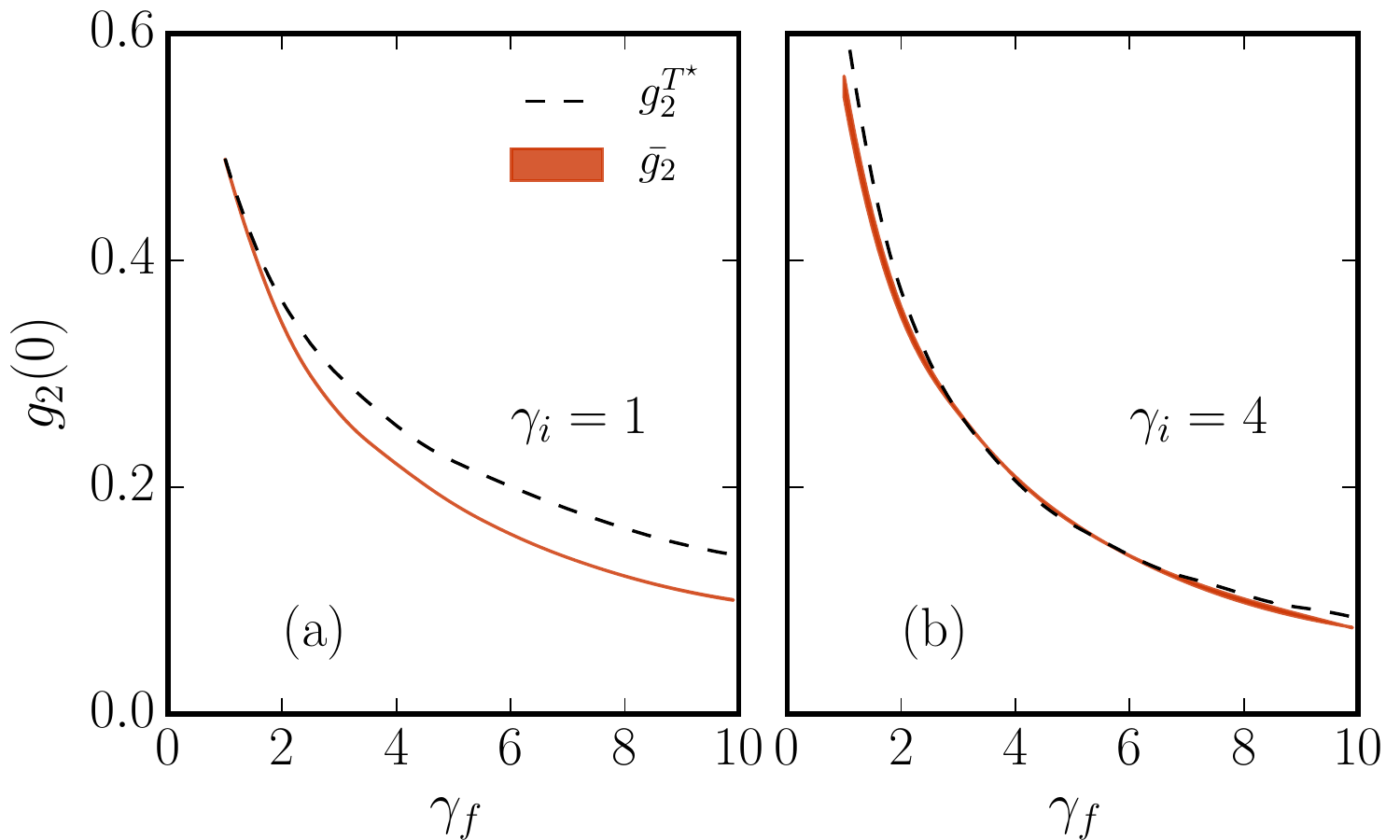}\caption{Time-dependent expectation value of local two-body correlations after
a quantum quench from the interacting ground state at $\gamma_{i}=1$
(a), and $\gamma_{i}=4$ (b), long-term dynamical averages (red continuous
lines) are compared to thermal averages at the effective temperature
set by energy conservation (black dashed lines). Uncertainties on
the thermal averages of the order of lines width, and are larger for
small values of $\gamma_{f}$. \label{fig:dyn-thermal}}
\end{figure}

We now turn to quenches from interacting initial states ($\gamma_{i}\neq0$)
to different interacting final states for which no results have been
obtained by means of exact BA nor simplified quench action method
so far. In Fig.~\ref{fig:dyn-thermal} we show the asymptotic equilibrium
values obtained with our t-VMC approach for quantum quenches from
$\gamma_{i}=1$ (Left panel), and $\gamma_{i}=4$ (Right panel) to
several values of $\gamma_{f}$. Since, by the variational theorem,
the ground state of $\mathcal{H}_{i}$ gives an upper bound for the
ground state energy of $\mathcal{H}_{f}$, the system is pushed into
a linear combination of excited states of the final hamiltonian. For
systems able to thermalize to the Boltzmann ensemble (BE), relaxation
to a stationary state described by the density matrix $\rho_{T^{\star}}=e^{-\mathcal{H}_{f}/T^{\star}}$,
at an effective temperature $T^{\star}$, would occur. Comparing the
stationary value, $\bar{g_{2}}(0)$, of our t-VMC calculations at
long times to the thermal values of the pair correlation functions,
$g_{2}^{T^{\star}}(0)$, a necessary condition for simple Boltzmann
thermalization is given by $\bar{g_{2}}=g_{2}^{T^{\star}}$. The effective
temperature, $T^{\star}$, is determined by imposing the energy expectation
value of the final Hamiltonian, $\mathcal{H}_{f}$, in the ground
state, $\Phi_{0}(\gamma_{i})$ of the initial Hamiltonian, $\langle\mathcal{H}_{\textrm{f}}\rangle_{T^{\star}}=\left\langle \Phi{}_{0}(\gamma_{i})\right|\mathcal{H}_{\text{f}}\left|\Phi{}_{0}(\gamma_{i})\right\rangle $.
Here, the thermal expectation value, $\langle\mathcal{H}_{\textrm{f}}\rangle_{T^{\star}}$
at the equilibrium temperature $T^{\star}$ is computed from the Yang-Yang
BA equations \cite{yang_thermodynamics_1969}. The quantity $\langle\mathcal{H}_{\textrm{f}}\rangle_{T^{\star}}$
then depends on a single parameter $T^{\star}$, that is fitted to
match the value of $\left\langle \Phi{}_{0}(\gamma_{i})\right|\mathcal{H}_{\text{f}}\left|\Phi{}_{0}(\gamma_{i})\right\rangle $.

As shown in Fig.\ref{fig:dyn-BA-comp}-(b), Boltzmann thermalization
certainly does not occur in the case for the Lieb-Liniger model when
quenching from a non-interacting state, $\gamma_{i}=0$, where we
find $\bar{g_{2}}\neq g_{2}^{T^{\star}}$. This can be understood
in terms of the existence of dynamically conserved charges (beyond
energy and density conservation) which can yield an equilibrium value
substantially different from the BE prediction. In particular, it
is widely believe that the Generalized Gibbs Ensemble (GGE) is the
correct thermal distribution approached after the quench \cite{rigol_fundamental_2016,kollar_relaxation_2008}.
Several constructive approaches for the GGE have been put forward
in past years \cite{caux2013timeevolution,caux2012constructing,nardis_relaxation_2015},
and the quench action predictions reported in Fig.\ref{fig:dyn-BA-comp}-(b)
converge to the GGE predictions for the thermal values. In Fig.\ref{fig:dyn-BA-comp}-(b)
we also show the thermal GGE values $g_{2}^{\text{GGE}}$ (rightmost
dashed lines), and notice that our results are much closer to the
GGE predictions than the simple BE. Deviations from the asymptotic
GGE results are observed at large $\gamma_{f}$, a regime in which
the accuracy of our approach is still sufficient to resolve the difference
between the BE and the long-term equilibration value.

For correlated initial ground states, $\gamma_{i}\neq0$, GGE predictions
are fundamentally harder to obtain than for the non-interacting initial
states, and the BE is the only reference thermal distribution we can
compare with at this stage. From our results we observe that the difference
between $\bar{g_{2}}$ and the simple BE prediction, $g_{2}^{T^{\star}}$,
is quantitatively reduced, see Fig. \ref{fig:dyn-thermal}. In particular,
for $\gamma_{i}=4$, the stationary values $\bar{g_{2}}$ are quantitatively
close to the ones predicted by the Boltzmann thermal distribution
at the effective temperature $T^{\star}$. Even though this quantitative
agreement is likely to be coincidental, the regimes of parameters
quenches studied here provide guidance for future experimental studies.
In particular, it will be of great interest to understand whether
a cross-over from a strongly non-Boltzmann to a close-to-Boltzmann
thermal behavior might occur as a function of the initial interaction
strength also for other local observables.

\section{Conclusions}

In this paper we have introduced a novel approach to the dynamics
of strongly-correlated quantum systems in continuous space. Our method
is based on correlated many-body wave-function systematically expanded
in terms of reduced $m$-body Jastrow functions. The unitary dynamics
in the subspace of these correlated states was realized using time-dependent
variational Monte Carlo. We have demonstrated the possibility or performing
calculations up to the three-body level, $m\le3$, for the Lieb-Liniger
model, for both static and dynamical properties. The improvement from
$m=2$ to $m=3$ provides an internal criterium to judge the validity
of our results whenever exact results are unavailable. Benchmarking
t-VMC with exact or numerical approaches whenever available, we have
found a very good agreement with existing results. For static properties,
our approach is at the level of state-of-the-art MPS techniques in
lattice systems and of latest c-MPS results for interacting gases.
For dynamical properties, we have investigated for the first time
general interaction quenches which are at the moment unaccessible
to Bethe-ansatz approaches. Since the general structure of our t-VMC
method does not depend on the dimensionality of the system, it can
be directly applied to bosonic systems in higher dimensions with a
polynomial increase in computational cost. The methods presented here
therefore pave the way to accurate out-of-equilibrium dynamics of
two-\ and three-dimensional quantum gases and fluids beyond mean
field approximations. 
\begin{acknowledgments}
We acknowledge discussions with J. De Nardis, M. Dolfi, M. Fagotti,
T. Osborne, and M. Troyer. We thank M. Ganahl for providing us the c-MPS results in Fig. \ref{fig:GS-prop}, 
J. Zill for the Bethe ansatz results in Fig. \ref{fig:dyn-BA-comp} (a), and J. De Nardis for the quench actions results in 
Fig. \ref{fig:dyn-BA-comp} (b).
This research was supported by the Marie Curie IEF program (FP7/2007-2013
- Grant Agreement No. 327143), the European Research Council Starting
Grant \textquotedbl{}ALoGlaDis\textquotedbl{} (FP7/2007-2013 Grant
Agreement No.~256294) and Advanced Grant \textquotedbl{}SIMCOFE\textquotedbl{}
(FP7/2007-2013 Grant Agreement No.~290464), the European Commission
FET-Proactive QUIC (H2020 grant No.~641122), the French ANR-16-CE30-0023-03
(THERMOLOC), and the Swiss National Science Foundation through NCCR
QSIT. It was performed using HPC resources from GENCI-CCRT/CINES (Grant
c2015056853). 
\end{acknowledgments}

\appendix

\section{Functional Structure of Many-Body Terms\label{sec:Functional-Structure-of}}

The local residuals $R^{(M)}(\mathbf{X};t)=i\partial_{t}U^{(M)}(\mathbf{X};t)-E_{\text{loc}}^{(M)}(\mathbf{X};t)$
are vanishing if the Schrödinger equation is exactly satisfied by
the many-body wave-function truncated at some order $M$. The local
energy $E_{\text{loc}}^{(M)}(\mathbf{X},t)$, however, may contain
effective interaction terms involving a number of bodies larger than
$M$, which leads to a systematic error in the truncation. However,
the structure of these additional terms stemming from the local energy
can be systematically used to deduce the functional structure of the
higher order terms. For example, the one-body truncated local energy
reads, for one-dimensional particles, 
\begin{eqnarray}
E_{\text{loc}}^{(1)}(\mathbf{X};t) & = & -\frac{1}{2}\sum_{i}\left\{ \left[\partial_{x_{i}}u_{1}(x_{i};t)\right]^{2}+\partial_{x_{i}}^{2}u_{1}(x_{i};t)\right\} +\nonumber \\
 &  & +\sum_{i}v_{1}(x_{i})+\frac{1}{2}\sum_{i\neq j}^{N}v_{2}(x_{i},x_{j}),\label{eq:Eloc1}
\end{eqnarray}
and contains a $2$-body term which cannot be accounted for exactly
by $u_{1}$. Introduction of a symmetric two-body Jastrow factor $u_{2}(x_{i},x_{j};t)$,
then leads to 
\begin{eqnarray}
E_{\text{loc}}^{(2)}(\mathbf{X};t) & = & E_{\text{loc}}^{(1)}(\mathbf{X};t)+\nonumber \\
 &  & -\frac{1}{2}\sum_{i\neq j}\left[\partial_{x_{i}}u_{1}(x_{i};t)\partial_{x_{i}}u_{2}(x_{i},x_{j};t)\right]+\nonumber \\
 &  & -\frac{1}{2}\sum_{i\neq j}\partial_{x_{i}}^{2}u_{2}(x_{i},x_{j};t)+\nonumber \\
 &  & -\frac{1}{2}\sum_{i\neq j}\sum_{k\neq i}\partial_{x_{i}}u_{2}(x_{i},x_{j};t)\partial_{x_{i}}u_{2}(x_{i},x_{k};t).\label{eq:Eloc2}
\end{eqnarray}
In the latter expression, one recognizes an effective two-body term
which can be accounted for by $u_{2}$ and an additional three-body
term in the form of product of two-body functions. The functional
form of the three-body Jastrow can be therefore deduced from this
additional term and formed accordingly: 
\begin{eqnarray}
u_{3}(x_{i},x_{j},x_{k};t) & = & \bar{u}_{3}(x_{i},x_{j};t)\bar{u}_{3}(x_{j},x_{k};t),\label{eq:jthree}
\end{eqnarray}
with two-body functions $\bar{u}_{3}(x_{i},x_{j};t)$ containing new
variational parameters to be determined. Upon pursuing this approach,
the expansion can be systematically pushed to higher orders and the
functional structure of the higher order functions inferred. The same
constructive approach we have discussed here is also valid for the
Schrödinger equation in imaginary-time $\partial_{\tau}U(\mathbf{X};\tau)=-E_{\text{loc}}(\mathbf{x};\tau)$,
and has been successfully used to infer the functional structure for
ground-state properties \cite{PhysRevE.68.046707}.

\section{Monte Carlo Sampling\label{sec:Monte-Carlo-Sampling}}

In order to solve the t-VMC equations of motion, Eq. (\ref{eq:eom}),
expectation values of some given operator $\mathcal{O}$ need to be
computed over the many-body wave-function $\Phi(\mathbf{X},t)$. This
is achieved by means of Monte Carlo sampling of the probability distribution
$\Pi(\mathbf{X})=|\Phi(\mathbf{X})|^{2}$ (In the following we omit
explicit reference to the time $t$, assuming that all expectation
values are taken over the wave-function at a given fixed time). An
efficient way of sampling the given probability distribution is to
devise a Markov chain of configurations $\mathbf{X}(1),\mathbf{X}(2),\dots\mathbf{X}(N_{c}-1),\mathbf{X}(N_{c})$
which are distributed according to $\Pi(\mathbf{X})$. Quantum expectation
values of a given operator $\mathcal{O}$ can then be obtained as
statistical expectation values over the Markov chain as 
\begin{eqnarray}
\frac{\left\langle \Phi\right|\mathcal{O}\left|\Phi\right\rangle }{\left\langle \Phi\right|\left.\Phi\right\rangle } & \simeq & \frac{1}{N_{c}}\sum_{i=1}^{N_{c}}\mathcal{O}_{\text{loc}}(\mathbf{X}(i)),\label{eq:avmonte}
\end{eqnarray}
where $\mathcal{O}_{\text{loc}}(\mathbf{X})=\frac{\left\langle \mathbf{X}\right|\mathcal{O}\left|\Phi\right\rangle }{\left\langle \mathbf{X}\right|\left.\Phi\right\rangle }$,
and the equivalence is achieved in the limit $N_{c}\rightarrow\infty$.

The Markov chain is realized by the Metropolis-Hastings algorithm.
Given the current state of the Markov chain, $\mathbf{X}(i)$, a configuration
$\mathbf{X^{\prime}}$ is generated according to a given transition
probability $T(\mathbf{X}(i)\rightarrow\mathbf{X}')$. The proposed
configuration is then accepted (i.e. $\mathbf{X}(i+1)=\mathbf{X}^{\prime}$
) with probability 
\begin{multline}
A(\mathbf{X}(i)\rightarrow\mathbf{X}')=\\
=\text{min}\left[1,\frac{\Pi(\mathbf{X}^{\prime})}{\Pi(\mathbf{X}(i))}\frac{T(\mathbf{X}^{\prime}\rightarrow\mathbf{X}(i))}{T(\mathbf{X}(i)\rightarrow\mathbf{X}')}\right],\label{eq:acceptmonte}
\end{multline}
otherwise it is rejected and $\mathbf{X}(i+1)=\mathbf{X}(i)$.

In the present t-VMC calculations we use simple transition probabilities
in which a single particle is displaced, while leaving all the other
particles positions unchanged. In particular, a particle index $p$
is chosen with uniform probability $1/N$ and the position of particle
$p$ is then displaced according to $x_{p}^{\prime}=x_{p}+\eta_{\Delta}$,
where $\eta_{\Delta}$ is a random number uniformly distributed in
$[-\frac{\Delta}{2},\frac{\Delta}{2}]$. The amplitude $\Delta$ is
an adjustable parameter and it can be typically chosen to be of the
order of the average inter-particle distance. With this choice, the
transition probability is simply 
\begin{eqnarray}
T(x_{p}\rightarrow x_{p}^{\prime}) & = & \frac{1}{N\Delta},\label{eq:transprob}
\end{eqnarray}
and the acceptance probability is therefore given by the mere ratio
of the probability distributions, $\frac{\Pi(\mathbf{X}^{\prime})}{\Pi(\mathbf{X}(i))}$.

\section{Finite Basis Projection\label{sec:Finite-Basis-Projection}}

The numerical solution of the equations of motion (\ref{eq:eom})
requires the projection of the Jastrow fields $u_{m}(\mathbf{r};t)$
onto a finite basis. The continuous variable $\mathbf{r}$ is reduced
to a finite set of $P$ values for each order $m$, $(m,\mathbf{r})\rightarrow(r_{1,m},r_{2,m}\dots r_{P,m})$.
We introduced a super-index $K$ spanning all possible values of the
discrete variables $r_{i,m}$. The complete set of variational parameters
resulting from the projection on the finite basis can then be written
as $u_{K}(t)$ and the associated functional derivatives read $\rho_{K}(t)$.

The integro-differential equations (\ref{eq:eom}) are then brought
to the algebraic form 
\begin{eqnarray}
\sum_{K^{\prime}}S_{K,K^{\prime}}\dot{u}_{K^{\prime}}(t) & = & -i\left\langle E_{\text{loc}}(t)\rho_{K}(t)\right\rangle ,\label{eq:eomproj}
\end{eqnarray}
where we have introduced the Hermitian correlation matrix 
\begin{eqnarray}
S_{K,K^{\prime}}(t) & = & \frac{\partial\left\langle \rho_{K}(t)\right\rangle _{t}}{\partial u_{K^{\prime}}}=\nonumber \\
 & = & \left\langle \rho_{K}(t)\rho_{K^{\prime}}(t)\right\rangle _{t}-\left\langle \rho_{K}(t)\right\rangle _{t}\left\langle \rho_{K^{\prime}}(t)\right\rangle _{t}.\label{eq:smatrix}
\end{eqnarray}
At a given time, all the expectations values in Eq. (\ref{eq:eomproj})
can be explicitly computed with the stochastic approach described
in Appendix \ref{sec:Monte-Carlo-Sampling}. We are therefore left
with a linear system in the $n_{\text{var}}$ unknowns $\dot{u}_{K}(t)$,
which needs to be solved at each time $t$.

In the presence of a large number of variational parameters, $n_{\text{var}}$,
the solution of the linear system can be achieved using iterative
solvers e.g., conjugate gradient methods, which do not need to explicitly
form the matrix $S$. Calling $n_{\text{iter}}$ the number of iterations
needed to obtain a solution for the linear system, the computational
cost to solve (\ref{eq:eomproj}) is $\mathcal{O}(M\times n_{\text{var}}\times n_{\text{iter}})$
as opposed to the $\mathcal{O}(M\times n_{\text{var}}^{2})$ operations
needed by a standard solver in which the matrix $S$ is formed explicitly.
In the present work we resort to the Minimal Residual (MinRes) method,
which is a variant of the Lanczos method, working in the Krylov subspace
spanned by the repeated action of the matrix $S$ onto an initial
vector. In typical applications we obtain that $n_{\text{iter}}\ll n_{\text{var}}$
and several thousands of variational parameters can be efficiently
treated. This is of fundamental importance when the continuous (infinite-basis)
limit must be taken, for which $n_{\text{var}}\rightarrow\infty$.

Once the unknowns $\dot{u}_{K}(t)$ are determined, we can solve numerically
the first-order differential equations given in Eq.~(\ref{eq:eom})
for given initial conditions $u_{K}(0)$. In the present work we have
adopted an adaptive 4th order Runge-Kutta scheme for the integration
of the differential equations.

\section{Lattice Regularization For The Lieb-Liniger Model\label{sec:Lattice-Regularization-For}}

We consider a general wave-function $\Psi(x_{1},x_{2}\dots x_{N})$
for $N$ one-dimensional particles, governed by the Lieb-Liniger Hamiltonian.
By means of Variational Monte Carlo, we want to sample $|\Phi(\mathbf{X})|^{2}$,
this is achieved via a lattice regularization, i.e. 
\begin{eqnarray*}
\int d\mathbf{X}|\Phi(\mathbf{X})|^{2} & \simeq & \sum_{l_{1},l_{2}\dots l_{N}}|\Phi(l_{1},l_{2}\dots l_{N})|^{2}
\end{eqnarray*}
where $l_{i}=\{0,a,\dots L-a\}$ are discrete particle positions,
$a$ the lattice spacing, $L$ the box size and $N_{s}=1+L/a$ the
number of lattice sites. As a discretized Hamiltonian we take 
\begin{multline*}
H_{a}\Phi(l_{1}\dots l_{N})=-\frac{\hbar^{2}}{2ma^{2}}\sum_{i}\left\{ \frac{4}{3}\left[\Phi(l_{1}\dots l_{i}-a,\dots l_{N})\right.+\right.\\
\left.\Phi(l_{1}\dots l_{i}+a,\dots l_{N})\right]-\\
\frac{1}{12}\left[\Phi(l_{1}\dots l_{i}-2a,\dots l_{N})+\Phi(l_{1}\dots l_{i}+2a,\dots l_{N})\right]+\\
\left.-\frac{5}{2}\Phi(l_{1}\dots l_{i},\dots l_{N})\right\} +\Phi(l_{1}\dots l_{N})\frac{g}{a}\sum_{i<j}\delta(l_{i},l_{j})
\end{multline*}
The first terms constitute just the fourth-order approximation of
the laplacian via central finite differences, whereas the last term
corresponds the two-body delta interaction part.

With this discretization, a two-body Jastrow factor reads 
\begin{eqnarray*}
u_{2}(x_{ij};t) & = & u_{2}(l_{i},l_{j};t),
\end{eqnarray*}
where $u_{2}(a,b;t)$ is a time-dependent matrix of size $N_{s}\times N_{s}$
which, in 1D and in the presence of translational symmetry, depends
only on $\textrm{dist}(a-b)$, i.e. it has $N_{s}/2$ variational
parameters.

\begin{figure*}
\includegraphics[clip,width=1\columnwidth]{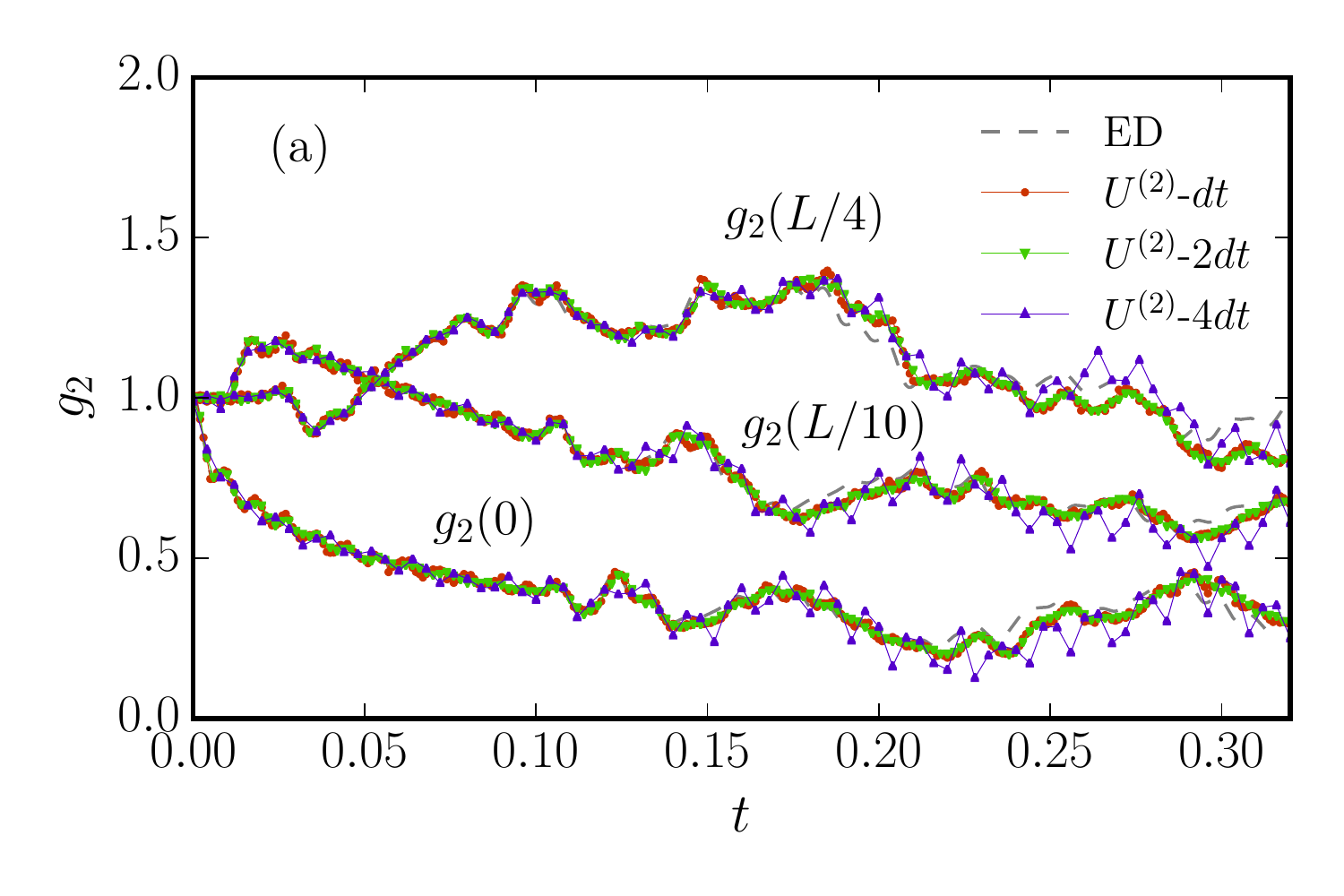}\includegraphics[clip,width=1\columnwidth]{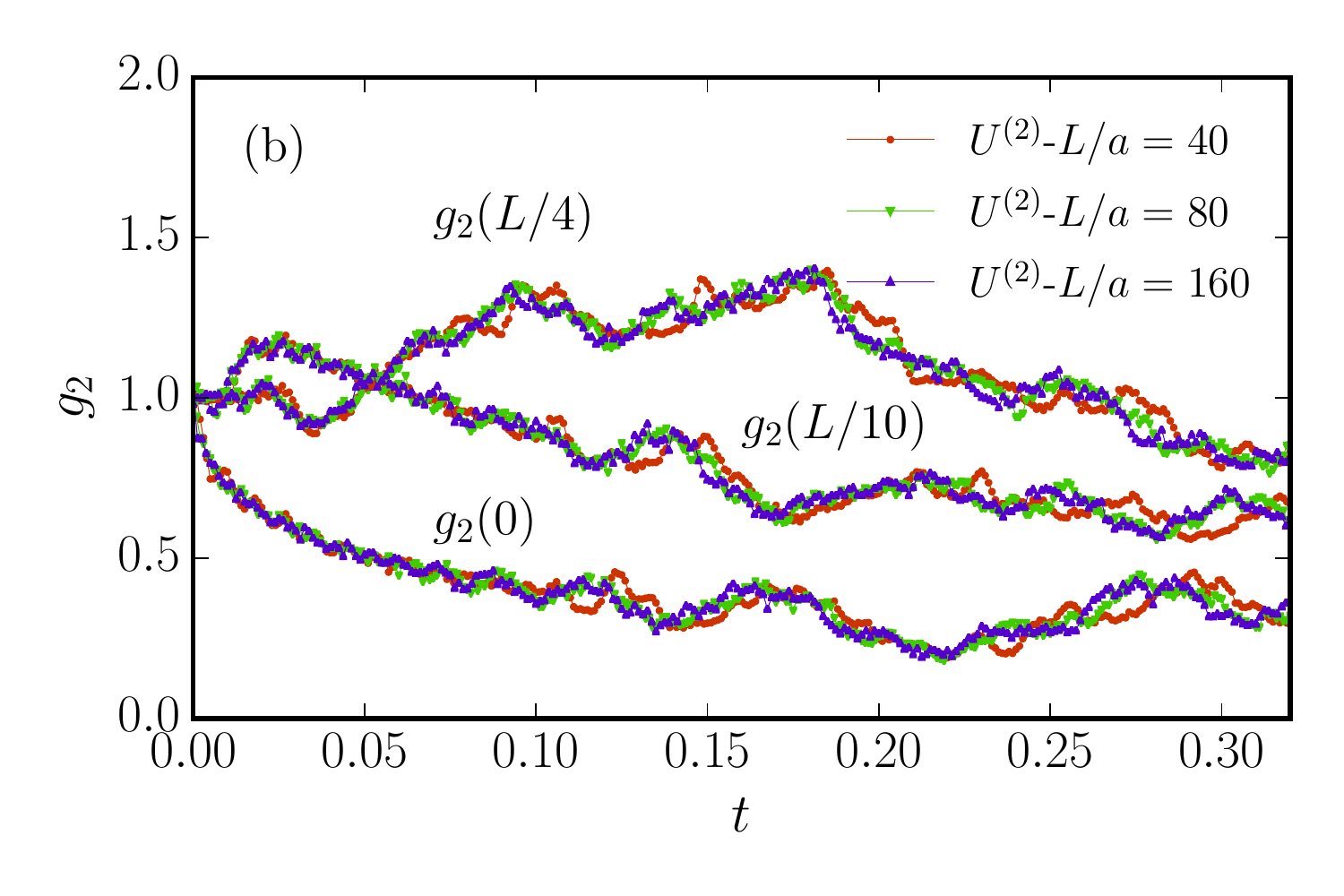}\caption{Time-dependent expectation value of the two-body correlations after
a quantum quench from a non-interacting state, $\gamma_{i}=0$, to
$\gamma_{f}=4$ at three different distance $|x_{1}-x_{2}|=0,L/10,L/4$.
Here, the system is on a lattice with $L/a=40$ lattice sites, the
full line is obtained by exact diagonalization of the Hamiltonian,
the other curves are from tVMC truncated at the level of $U^{(2)}$.
In the left figure (a), we show the convergence with different time
step discretization. On the right figure (b), we show the approach
to the continuum for t-VMC simulations using $U^{(2)}$ for discretizations
$L/a=40$, $80$ and $160$. \label{fig:dyn-ED-dt}}
\end{figure*}

\begin{figure*}
\includegraphics[clip,width=1\columnwidth]{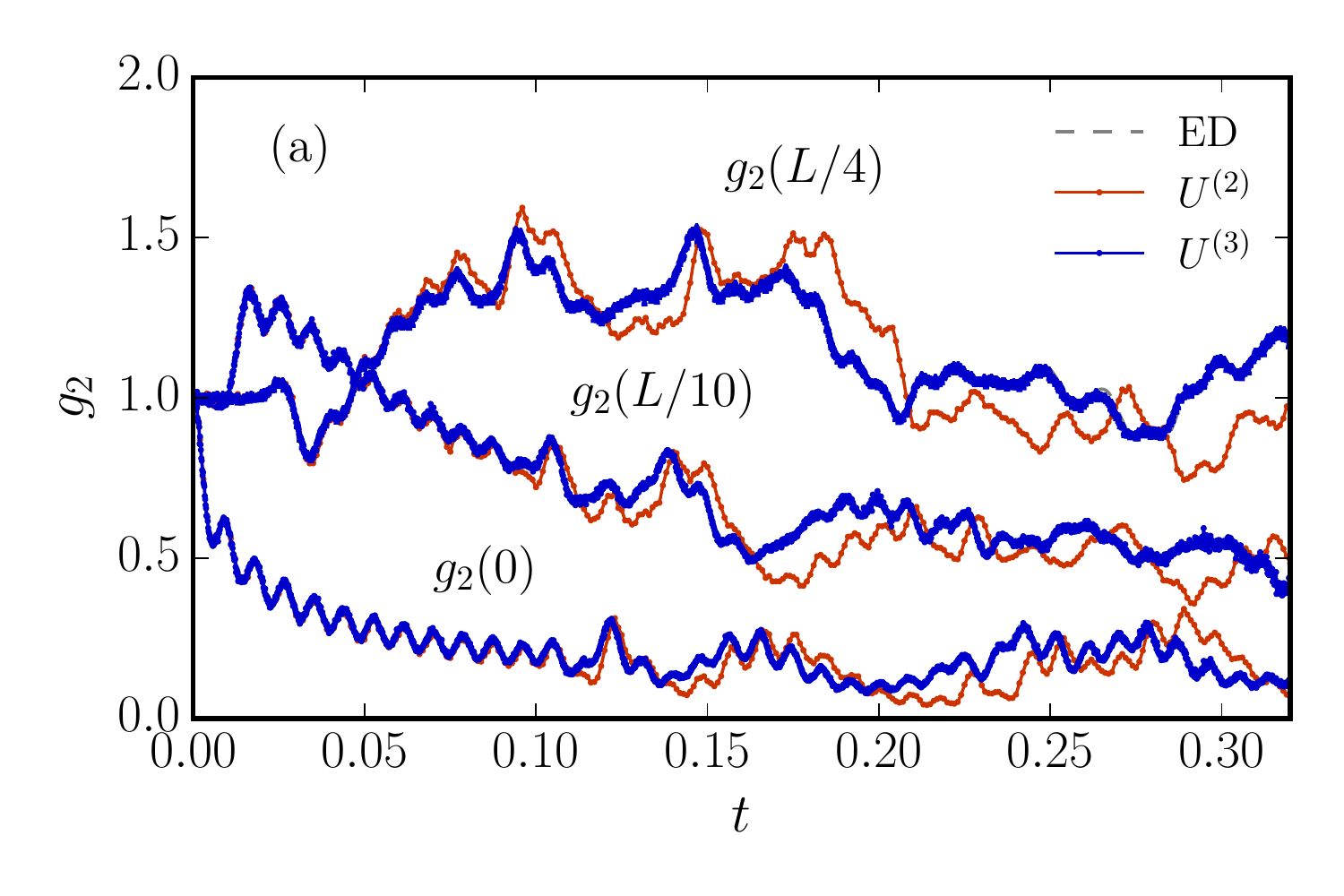}\includegraphics[clip,width=1\columnwidth]{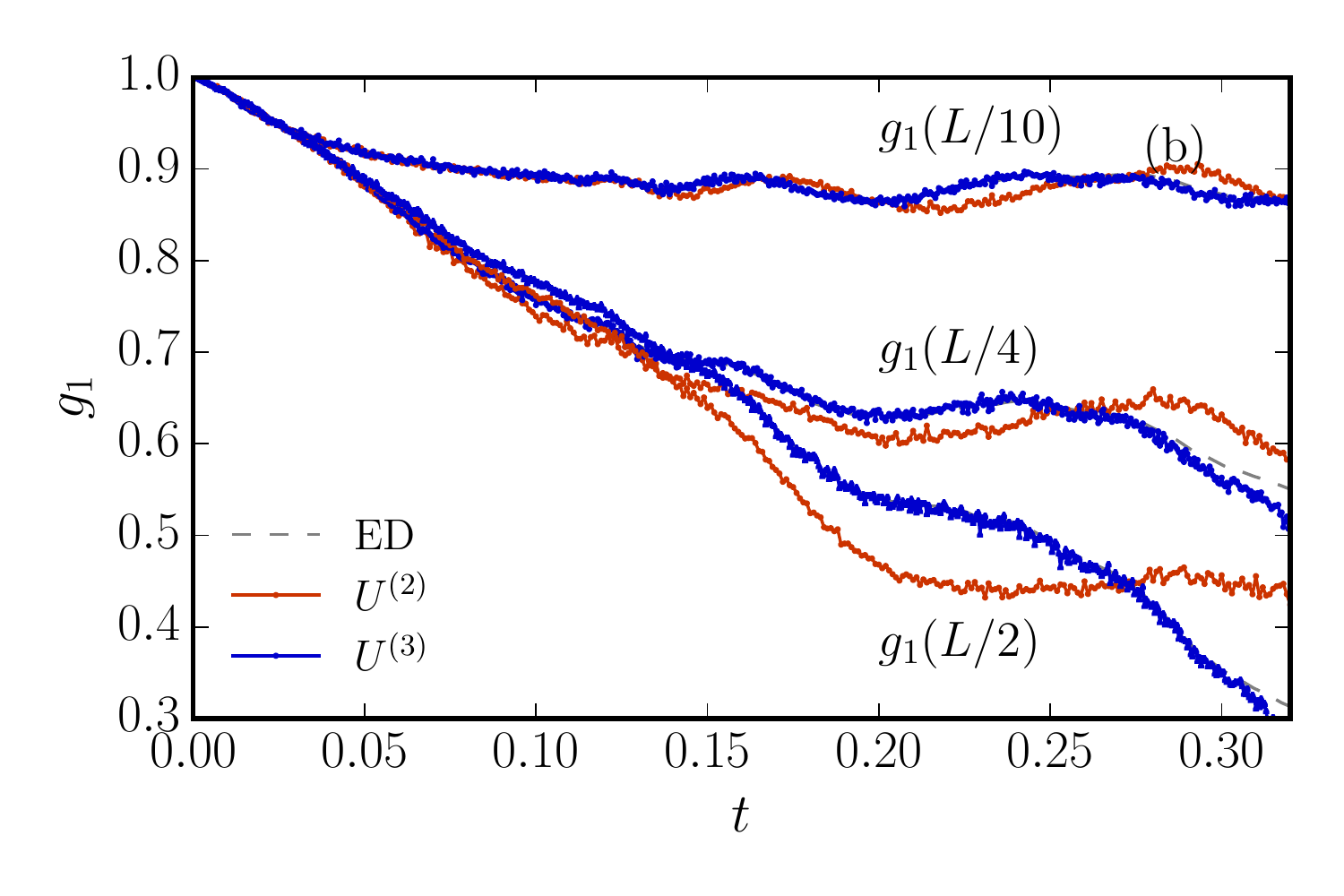}\caption{(a) Time-dependent expectation value of the two-body correlations,
$g_{2}(r,t)$, after a quantum quench from a non-interacting state,
$\gamma_{i}=0$, to $\gamma_{f}=8$ at three different distance $|x_{1}-x_{2}|=0,L/10,L/4$.
Here, the system is on a lattice with $L/a=40$ lattice sites, the
full line is obtained by exact diagonalization of the Hamiltonian,
the other curves are from t-VMC truncated at the level of $U^{(2)}$
or $U^{(3)}$. In contrast to $\gamma_{f}=4$ shown in Fig.\ref{fig:dyn-ED-dt},
systematic differences of $U^{(2)}$ compared to the exact results
are more visible here, the exact dynamics is recovered by inclusion
of three-body terms, $U^{(3)}$, into the tVMC wave function. (b)
shows the off-diagonal single particle density matrix, $g_{1}(r,t)$,
at three different distances, $r=L/10$, $L/4$ and $L/2$. \label{fig:dyn-ED-g1}}
\end{figure*}

\section{Benchmark Study for $N=3$ on a Lattice\label{sec:Lattice-Benchmark}}

Here, we use exact diagonalization of a Hamiltonian within a given
finite basis for a quantitative test of our method. Exact diagonalization
is limited to very small systems on a finite basis, and we haven chosen
a system containing $N=3$ particles on $L/a=40$ lattice sites as
a simple, but highly non-trivial reference. In contrast to our comparison
with BA methods, all observables can be accessed by exact diagonalization
and we have used the off-diagonal one body density matrix $g_{1}(r,t)$
and the pair correlation function $g_{2}(r,t)$ at different distances
$r=|x_{1}-x_{2}|$ of two particles after time $t$ where the system
is quenched from the non-interacting initial state, $\gamma_{i}=0$,
to a final interaction $\gamma_{f}>0$, to provide a benchmark on
a more general observable.

We first benchmark the influence of the time-step lattice size discretization
$\Delta t$ error on $g_{2}$. From Fig.\ref{fig:dyn-ED-dt}(a) we
see that the t-VMC dynamics is stable over a long time and the time
step error can be brought to convergence. Further, we see that for
final interaction $\gamma_{f}=4$ the truncation at the two-body level,
$U^{(2)}$, introduces only a small systematic error, mainly a dephasing
effect, which is almost negligible at the scale of the figure. Due
to the stochastic noise of the Monte Carlo integration, t-VMC introduces
additional high frequency oscillations which are, however, well separable
from the deterministic propagation. The amplitude of these high frequency
oscillations also quantifies the purely statistical error of our data.

Whereas exact diagonalization is limited to rather small basis sets,
we can access much a larger basis within t-VMC. In Fig.\ref{fig:dyn-ED-dt}
(b) we show results within the $U^{(2)}$ approximation with $L/a=80$
and $L/a=160$ with time discretization $\Delta t\sim(L/a)^{2}$.
We see that the basis set truncation in general introduces a dephasing
at large enough time.

The systematic error of $U^{(2)}$ increases towards quenches to stronger
interaction strength and becomes more visible for $\gamma_{f}=8$
shown in Fig.\ref{fig:dyn-ED-g1}(a). However, even in this case,
the most important effect remains to be a simple dephasing, a small
shift of averaged quantities is probable, but difficult to quantify
precisely. Introducing a general three-body Jastrow fields, $U^{(3)}$,
described in detail in Appendix \ref{sec:3body}, the systematic error
for $N=3$ can be fully eliminated.

In figure \ref{fig:dyn-ED-g1} (b), we also benchmark the possibility
of calculating the off-diagonal one-body density matrix after a quench.
Here the systematic error of $U^{(2)}$ is more pronounced at smaller
$\gamma_{f}$ in the long range and time regime.

From our study of the three particle problem, we conclude that truncation
of the many-body wave function at the level of $U_{2}$ may provide
an excellent approximation for $g_{2}(r,t)$ for quenches involving
not too strong interaction strengths, $\gamma\lesssim5$. The systematic
error due to the $U^{(2)}$ truncation is mainly a dephasing at large
times involving small relative errors of time averaged quantities.
Similar systematic dephasing errors will occur for too large time
discretization or basis set truncation.

Since our method provides a parametrization of the full wave function
for a given time, many different observables can be evaluated via
usual Monte Carlo methods. However, the quality of different observables
may vary and depend more sensitive on the inclusion of higher order
correlations $U^{(n)}$ with $n>2$ as in the case of the single body
density matrix. Although these higher order terms are computationally
expensive, the scaling is not exponential, and we have explicitly
shown that calculations with $n=3$ are feasible. We notice that the
computational complexity may be further reduced by functional forms
adapted to the problem \citep{PhysRevB.74.104510}.

\section{General Structure of $U^{(3)}$\label{sec:3body}}

For a general time dependent wave function, we have to go beyond the
usual ground state structure of the three-body Jastrow given in Eq.~(\ref{eq:jthree}).
Here, we provide details of our three-body term in a general form
beyond the present application in one dimension.

Introducing $M$ basis functions, $b^{a}(r)$, $a=1,\dots M$, we
can introduce many-body vectors \citep{PhysRevB.74.104510}, $B_{i\alpha}^{a}=\sum_{j}{\bf r}_{ij}^{\alpha}b^{a}(r_{ij})$,
where $\alpha=1,\dots D$ indicates the summation over directions
and $i=1,\dots N$. The variational parameters of a general three-body
structure can then be written in terms of a matrix $w_{ab}$, such
that 
\begin{equation}
\sum_{i\ne j\ne k}u_{3}({\bf r}_{1},{\bf r}_{2},{\bf r}_{3})=\sum_{ab}w_{ab}W^{ab},\quad W^{ab}=\sum_{i\alpha}B_{i\alpha}^{a}B_{i\alpha}^{b}
\end{equation}
In order to reduce the variational parameters ($\sim M^{2}$), we
may perform a singular value decomposition of the matrix $w_{ab}$
to reduce the effective degrees of freedom.


\end{document}